# ON SECONDARY ATOMIZATION AND BLOCKAGE OF SURROGATE COUGH DROPLETS IN SINGLE AND MULTI-LAYER FACE MASKS

**Shubham Sharma[1], Roven Pinto[1], Abhishek Saha[2], Swetaprovo Chaudhuri[3], Saptarshi Basu[1]**

[1]Department of Mechanical Engineering, Indian Institute of Science, Bengaluru, India

[2]Department of Mechanical and Aerospace Engineering, University of California San Diego, La Jolla, USA

[3]Institute for Aerospace Studies, University of Toronto, Toronto, Canada

## Abstract

By now it is well-understood that the usage of facemasks provides protection from transmission of viral loads through exhalation and inhalation of respiratory droplets. Therefore, during the current Covid-19 pandemic the usage of face masks is strongly recommended by health officials. Although three-layer masks are generally advised for usage, many commonly available or homemade masks contain only single and double layers. In this study, we show through detailed physics based analyses and high speed imaging that high momentum cough droplets on impingement on single- and double-layer masks can lead to significant partial penetration and more importantly atomization into numerous much smaller daughter droplets, thereby increasing the total population of the aerosol, which can remain suspended for a longer time. The possibility of secondary atomization of high momentum cough droplets due to impingement, hydrodynamic focusing and extrusion through the microscale pores in the fibrous network of the mask has not been explored before. However, this unique mode of aerosol generation poses a finite risk of infection as shown in this work. We also demonstrate that in single layer masks close to 70 % of a given droplet volume is atomized and only 30 % is trapped within the fibers. The entrapped volume is close to 90 % for double layer masks which still allows some atomization into smaller droplets. We however found that a triple-layer surgical mask permits negligible penetration and hence should be effective in preventing disease transmission.

## Introduction

The transport of pathogen loaded respiratory droplets from an infected person can result in the spread of viral loads to a susceptible person triggering global pandemics, like the ongoing COVID- 19 [1–3]. Droplets are ejected by an infected person while breathing, talking, coughing, singing, spitting, or sneezing and can remain airborne for a long time, depending on its initial size and ambient conditions [4–7]. These aerosolized droplets, when containing viral loading, can further infect a healthy person by their ingestion through oral or nasal passages into the respiratory tracts [8]. The infection probability of the droplet nuclei or the fomite depends on their initial viral loads [5,9,10], and their endurance in different ambient conditions [7,11–13]. Thus, wearing a face mask and maintaining social distancing in public settings is advised by the scientific and medical community for restricting the spread of the disease through droplets [14,15].

In particular, face masks are specifically important in both arresting the respiratory droplets ejected from individuals during respiratory events and limiting their ingestion during breathing processes. Although all masks are in general effective in reducing these transmissions, the relative effectiveness depends on the type of mask used. Hui et al. [16] discussed the effectiveness of face masks in confining the distance traveled by aerosol dispersions during human coughs. They showed that the turbulent flow induced during coughs without a mask could traverse an average distance of 70 cm from its initial position, and the use of a surgical mask and N95 mask reduces this distance approximately by a factor of 2.3 and 4.5, respectively. Although the N95 mask effectively restricted the forward translation of cough puffs, the sidewise leakage was still evident for these masks. Fischer et al. [17] discussed a cost-effective optical measurement method for finding mask efficacy in filtering respiratory droplets generated during the human speech. The droplet counts and their rate of ejection were compared for different mask surfaces, and it was found that cotton masks have identical safety as surgical face masks, while alternatives like neck gaiters or bandanas provide minimal protection. Dbouk and Drikakis [18] have numerically shown that a few droplets are transmitted to longer distances even after being obstructed by the face mask, and the efficiency of a face mask keeps on diminishing with increasing cough cycles. Verma et al. [19] compared the efficiency of different commercially available face masks in obstructing respiratory jets. They used a laser sheet illuminating the aerosols and calculated the distance traveled by the jets for the case of the unmasked and masked subjects. The use of face-covering significantly reduced the distance traveled by the jets; however, a minimal amount of aerosol leakage was found from the sides of the face mask. A similar study was conducted by Kahler and Hain [20] for a much smaller size of suspended droplets (0.1 - 2 $\mu$m) which suggested the use of particle filtering units in masks for increasing their effectiveness.

The available literature indicates that N95 masks are effective in limiting the spread of dispersions during human coughs, but their shortage and high costs in the ongoing pandemic has forced policy makers to shift to other alternatives like single- or multi-layer surgical masks or other homemade substitutes. Previous work on the surgical face mask has been mainly concentrated on determining the spreading distance of cough puffs and their leakages from the sidewalls of the mask[14,16,18–23], and addresses only the smaller sized droplets (~ 0.1 to 100 μm) which can easily transmit through the porous network of the mask. These studies lack in presenting the evolution of the impinging droplet inside the face mask. In particular, the fluid dynamical aspects of cough droplets impinging on the mask which covers the droplet penetration criteria, atomization mechanism and the final size distribution of the daughter droplets, remain mostly unexplored.

In the present investigation, we have focused on these aspects and studied the breakup mechanism of large cough droplets impacting a single- or multi-layer surgical mask. It is noted that the respiratory events release a plethora of droplet sizes, spanning from submicron to few millimeters [4,24,25], with an average velocity of 10 m/s[23,24,26,27]. Although the large droplets when ejected without the restriction of masks, travel only a limited distance before settling on the ground and as such are considered to be less important in direct transmission, we will show that these large droplets may lead to the fragmentation and regeneration of numerous tiny daughter droplets with significant translational velocity in single-layer surgical mask. These small droplets can move to longer distances and can remain aerosolized in the medium for longer durations increasing the risk of infections.

This paper is organized as follows; first, we describe the experimental setup used, followed by the results of droplet impact on different layered masks. Next, a scaling analysis for the criteria of droplet penetration through mask is conducted and compared with experimental data. This is followed by a theoretical model for predicting the size distribution of atomized droplets, validated by experimental data. Next, a probability distribution of the daughter droplet sizes is plotted which shows that atomization through single- and double-layer masks results in a majority of the daughter droplets falling in the possible regime of aerosolization. Finally, we discuss the effectiveness of different masks in trapping virus emulating nanoparticles from the impacting droplet.

## Results

**Experiments**

As mentioned before, the purpose of this work is to simulate impact of large cough droplets on single-, double-, and triple-layer surgical masks and thereby, evaluate their relative efficacy in restricting these droplets. Figure 1(a) shows a general schematic of these impact events during actual usage where droplets ejected during human coughs land on masks used as face covering. A zoomed-in view shows the droplet impacting on the inner layer of a single-layer or double layer mask and disintegrating into finer daughter droplets on the other side of the mask. Now to simulate a cough event in experiments, a piezo-actuated droplet dispenser (Nordson pico-pulse) is used, which ejects a DI water droplet of ~ 620 $\mu$m size with an impact velocity of ~10 m/s. A set of at least 10 experimental runs is done for each experimental case. It will be shown later secondary atomization as shown in Fig. 1a happens only for single and double layer masks for a range of impact velocities. Single- and double-layer masks are effective in blocking droplets with low momentum especially during talking and breathing. A high-speed shadowgraphy setup (see Fig. 1(b)) consisting of a laser source and a high-speed camera is used for visualizing the single droplet impingement on the masked surface (for further details, see "Methods" section).

Surgical masks from two different companies (locally supplied (mask A) and Novel mask Aavanzr Pharmaceuticals Pvt. Ltd. (mask B)) and with different numbers of protection layers (single, double, triple) are used during the experiments. The Scanning Electron Microscopy (SEM) images of mask A are shown in Fig. 1(c, d) for single- and double-layer masks, respectively. These images depict the porous network formed by the threads of the mask layer. A similar structure is found in mask B (see Fig. 1(f, g)). A single mask layer has a range of pore diameters, and the average effective pore diameter was found to be $\sim$ 30 $\mu$m for both mask A and mask B. For the double- and triple-layer masks, it is derived to be $\sim$ 17 $\mu$m and $\sim$ 12 $\mu$m, respectively (see Supplementary Fig. S1). Thus, overlapping layers of these masks reduce the effective porosity. The mask material used for a single-layer is hydrophobic (Fig. 1(e, h)), and the contact angle is measured to be $123 \pm 4^\circ$ and $115 \pm 8^\circ$ for mask A and mask B, respectively.

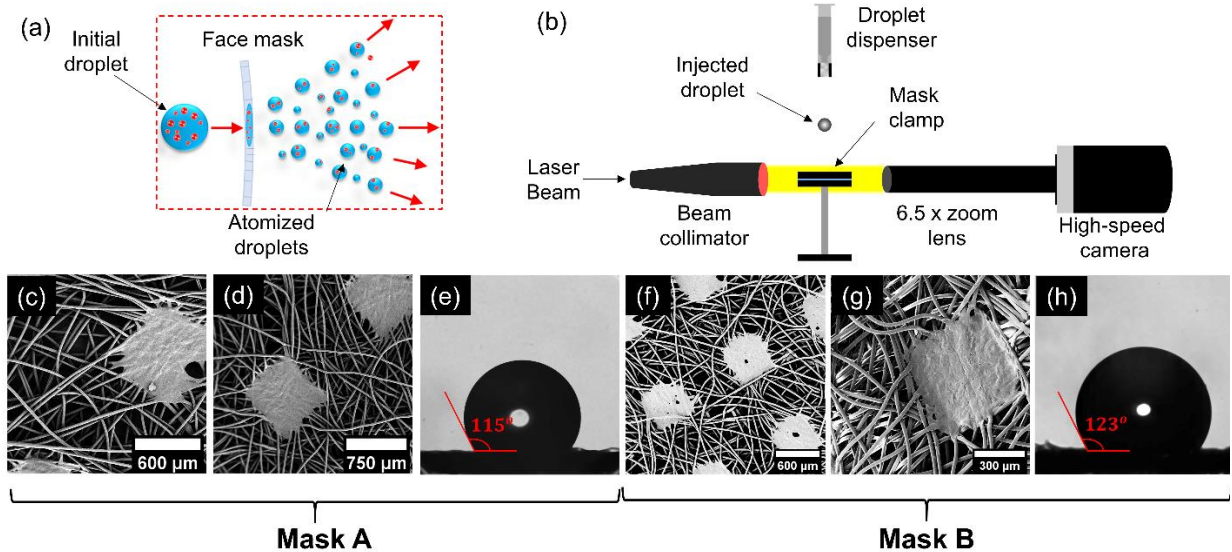

**Figure 1 Droplet atomization through face mask.** (a) Schematic diagram of droplets ejected during human cough. The larger droplet with high momentum get atomized into numerous tiny droplets after impacting on single and double layer mask surface. Note triple layer mask surface does not lead to any atomization (b) High-speed shadowgraphy imaging setup capturing the breakup dynamics of the impacting droplet. (c, f) SEM images showing the variable pore size in a single-layer of mask A and mask B, respectively. (d, g) SEM images showing the variable pore size in a double-layer of mask A and mask B, respectively. (e, h) The contact angle of a droplet on the surfaces of mask A and mask B, respectively.

**Droplet impact on different layered masks**

The time-sequence images of a droplet impacting on different layered mask A are shown in Fig. 2. The reference time is taken from the instance of the droplet impact on the mask layer. The impacting droplet has an initial diameter ($D_i$) of 617.70 ± 24 $\mu$m and impact velocity ($v_i$) of 10.12 ± 0.43 m/s. Figure 2(a) shows the case of a single-layer mask, in which the impacting droplet is fragmented into multiple liquid ligaments (see t = 100 - 450 ms), and these ligaments subsequently undergo secondary atomization into multiple daughter droplets (see t = 450 - 950 $\mu$s). Figure 2(b) shows the case of a double-layer mask in which the number of droplets penetrating through the mask are significantly less compared to a single-layer mask (see t = 250 - 1150 $\mu$s) due to a reduction in the effective porosity. The cylindrical ligaments are not prominent in this case due to the presence of the second mask layer. Apart from surgical masks, few locally sourced cloth masks with single and double layers are also investigated, and results are similar to the surgical mask of the corresponding number of layers (see Supplementary Fig. S3). Figure 2(c) shows the droplet impacting on the triple-layer surgical mask. Due to the much smaller effective porosity of triple-layer masks, half of the total experimental runs produced no droplet penetration and the remaining runs resulted in the penetration of only a single droplet (see Fig. 2(c) at t = 600 - 1000 μs) through the mask A. Mask B also shows similar results (see Supplementary Fig. S2), but no penetration was observed in a triple-layer of mask B. Similarly, no penetration is observed for an N95 face mask. Thus, triple-layer masks and N95 masks are not only useful in restricting larger respiratory droplets, but they also inhibit the further atomization of droplets that are ejected during the cough of an infected person.

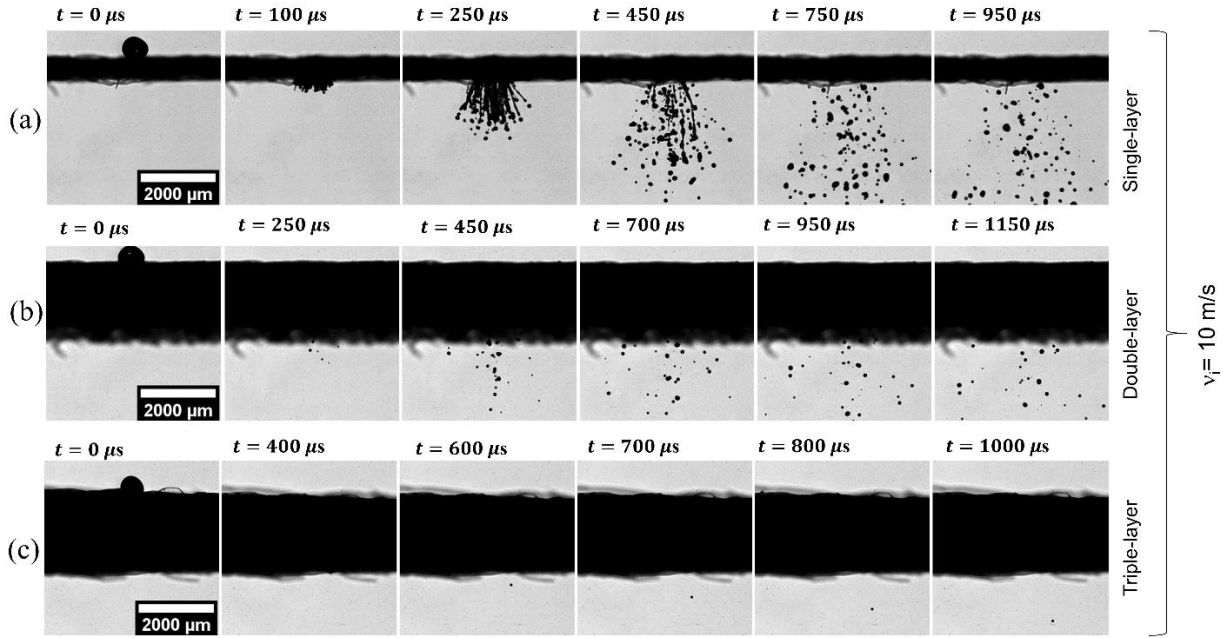

**Figure 2 Dynamic images of a droplet impacting on different layered mask A.** The water droplet impacting on the mask surface has a "$We \sim 880$" and is recorded at 20000 fps. (a - c) Time sequence images of droplet impingement on a single-, double-, and triple-layer mask, respectively. The total number count of atomized droplets is significantly higher for the single-layer mask in comparison with the double-layer mask, while only a single droplet penetrates through the triple-layer mask (see Fig. 2(c) at "t = 600 - 1000 μs"). Similar results are obtained for mask B (see Supplementary Fig. S2). Scale bar description is included in the figures.

**Droplet penetration criteria**

Sahu et al.[28] found that for any fiber-liquid combination, there exists a threshold impact velocity above which liquid can penetrate the porous network of fibers irrespective of its hydrophobicity. The scaling analysis[28] for determining the criteria of droplet penetration is obtained as follows. Since the surface tension effect can be neglected, the initial kinetic energy $\left(E_k \approx \rho_w \left(\frac{v_i D_i}{\varepsilon}\right)^2 D_i^3\right)$ of the penetrating droplet can be considered to be lost into dissipation energy $\left(E_d \approx \frac{\mu_w}{\varepsilon} \left(\frac{v_i D_i}{\varepsilon}\right) \varepsilon D_i \left(\frac{D_i}{\varepsilon}\right)^3 t_m\right)$ as the liquid ligaments pass through the porous network of the mask. Here, $\varepsilon$ and $t_m$ are the pore size and thickness of the mask layer, $\rho_w$ and $\mu_w$ are the density and dynamic viscosity of the droplet, respectively. If the kinetic energy overcomes the dissipation energy, the impacting droplet penetrates through to the other side of the mask layer. The ratio of two energies, thus, gives the scaling argument for this criterion:

$$\frac{E_k}{E_d} \approx \rho_w \left(\frac{v_i D_i}{\varepsilon}\right)^2 D_i^3 \cdot \frac{\varepsilon^5}{\mu_w v_i D_i^4 \varepsilon t_m} \approx Re_\varepsilon \left(\frac{\varepsilon}{t_m}\right) \qquad (1)$$

Here $Re_\varepsilon = \left(\frac{\rho_w v_i \varepsilon}{\mu_w}\right)$ is Reynolds number based on the pore size and droplet impact velocity. Therefore, for the droplet to penetrate through the mask layer, the above ratio should be much greater than one $\left(\frac{E_k}{E_d} \gg 1\right)$, and the droplet penetration criteria is obtained as[28]:

$$Re_\varepsilon \left(\frac{\varepsilon}{t_m}\right) \gg 1 \tag{2}$$

Thus, the droplet penetration depends on the impact velocity, thickness and pore size of the mask, while it is independent of the diameter of impacting droplet [28]. For validating the applicability of this criteria, experiments are conducted with different droplet impact velocities of 2 - 10 m/s and keeping same droplet diameters for all cases. The magnitudes of left and right terms of Eq. (2) for different cases are shown in Table 1. In single- and double-layer masks, for a droplet impact velocity of 10 m/s, $Re_\varepsilon (\varepsilon/t_m) \gg 1$ therefore, the impacting droplet can penetrate through the mask layer (see Fig. 2a, b). Whereas for a triple-layer mask $Re_\varepsilon (\varepsilon/t_m) \sim 1$, thus, no, or minimal penetration is observed for this case (see Fig. 2c). At an impact velocity of 7.5 m/s, the penetration criteria is satisfied for a single- and double-layer mask and hence we observed droplet penetration for these cases, and no penetration is observed for a triple-layer mask (see Fig. 3(a-c)). At lower impact velocity of 3 and 5 m/s on a single-layer mask, $Re_\varepsilon (\varepsilon/t_m) > 1$ thus we obtained droplet penetration for this case while no penetration was observed for multiple layer masks (see Fig. 3(d-f)). At a much lower impact velocity of 2 m/s, no droplet penetration was observed even through a single-layer mask (see Fig. 3(g)). A qualitative look at Eq. 2 shows that due to smaller pore size, the Reynolds number decreases, and the thickness to pore ratio increases for a multi-layer mask. Thus, the droplet penetration criteria given by Eq. 2 is no longer satisfied for these masks, and we observed a minimal and no droplet penetration for double- and triple-layer face masks, respectively. The detailed mechanism of droplet breakup and the size distribution of ejected droplets is discussed in the following section.

|  | Single-layer | Double-layer | Triple-layer |
|---|---|---|---|
| $Re_\varepsilon\ (\varepsilon/t_m)$ at $v_i$ = 10 m/s | 27.22 | 4.37 | 1.45 |
| $Re_\varepsilon\ (\varepsilon/t_m)$ at $v_i$ = 7.5 m/s | 20.42 | 3.28 | 1.09 |
| $Re_\varepsilon\ (\varepsilon/t_m)$ at $v_i$ = 5 m/s | 13.61 | 2.19 | 0.73 |
| $Re_\varepsilon\ (\varepsilon/t_m)$ at $v_i$ = 3 m/s | 8.17 | 1.31 | 0.44 |
| $Re_\varepsilon\ (\varepsilon/t_m)$ at $v_i$ = 2 m/s | 5.45 | 0.87 | 0.29 |

**Table 1: Penetration criteria for different layered mask A for different impact velocities and droplet size of ~ 620 μm.** The red and green regions indicate the cases of droplet penetration and no penetration through the mask, respectively.

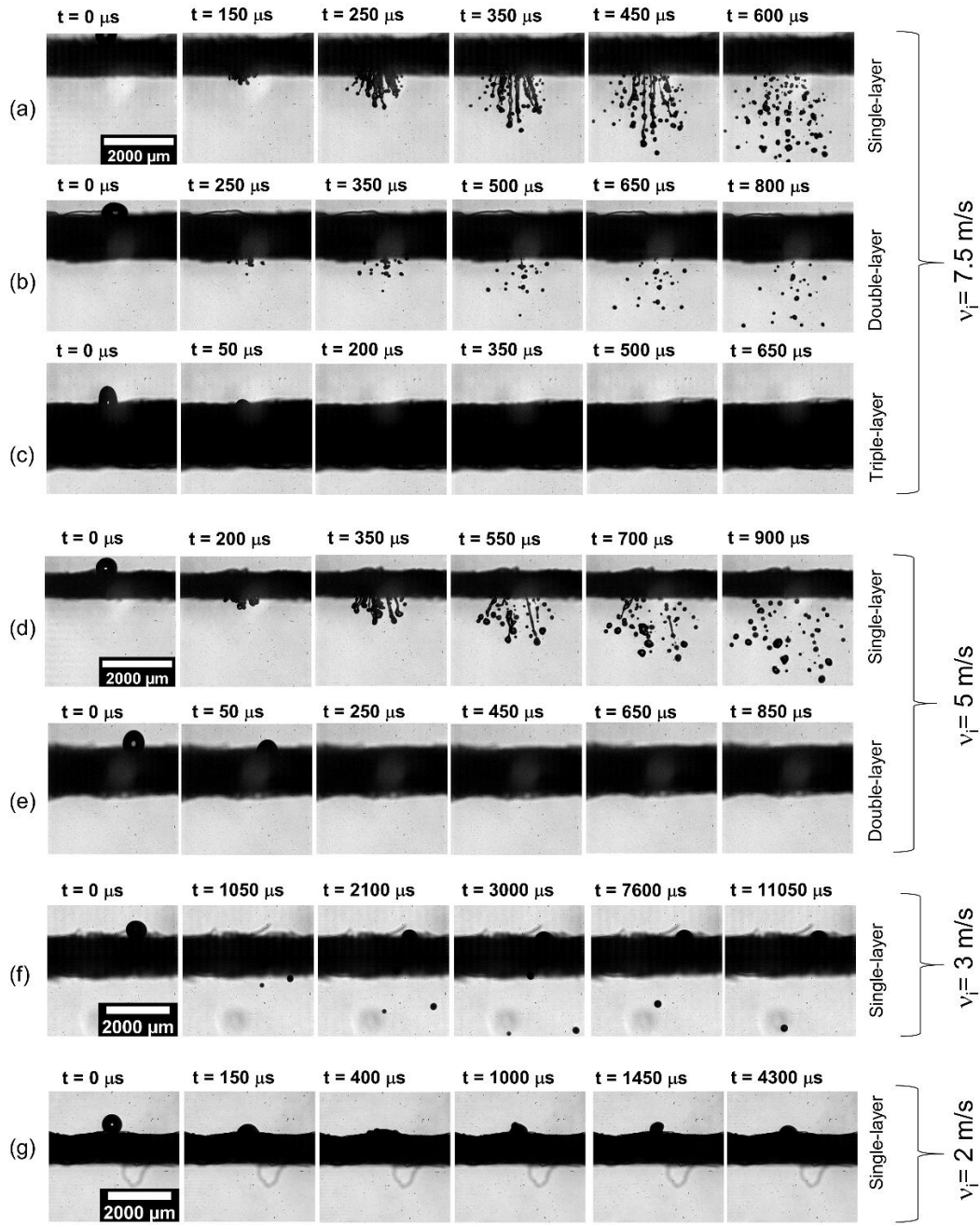

**Figure 3 Dynamic images of a droplet impacting at different impact velocities.** (a-c) Time sequence images of droplet impinging at "$v_i$ = 7.5 m/s" on a single-, double- and triple-layer of mask A, respectively. Droplet penetration is observed for single- and double-layer masks while no penetration is observed for triple layer mask. (d, e) Time sequence images of droplet impinging at "$v_i$ = 5 m/s" on a single- and double-layer of mask A, respectively. Droplet penetration is only observed for single-layer masks while no penetration is observed for double- and triple-layer (not shown in figure) mask. (f, g) Time sequence images of droplet impinging at "$v_i$ = 3 m/s and 2m/s" on a single layer of mask A, respectively. Impacting droplet penetrates through the single-layer mask at "$v_i$ = 3 m/s" while no penetration is observed for "$v_i$ =2 m/s". Scale bar description is included in the figure.

**Droplet atomization mechanism**

The zoomed-in images of droplet atomization are shown in Fig. 4. For impacting droplets, Weber number $\left(We = \frac{F_{inertia}}{F_{surface\ tension}} = \frac{\rho_w v_i^2 D_i}{\sigma}\right)$ governs the relative importance of inertia and surface tension forces acting on the droplet, where $\sigma$ is the surface tension of water (droplet liquid) in the air (surrounding gas medium). For a higher $We$ as in the present case ($We$ = 875.9) the inertial forces dominate over the surface tension forces, resulting in significant deformation of the droplet as it impacts the mask surface. Based on the penetration criteria discussed in the previous section, the impacting droplet extrudes through a single-layer mask in the form of cylindrical ligaments (see Fig. 4(a) at t = 50 μs). The length of these ligaments increases over time (see Fig. 4(a) at t = 50 - 250 μs) due to which instabilities in the form of capillary waves are formed on its surface. Among all the instabilities, few dominant unstable wavelength amplitudes grow over time and result in thinner and thicker diameter regions along the length of the ligament (see Fig. 4(a) at t = 300 μs). Hence, different surface curvatures are formed on the ligament, which results in a Laplace pressure gradient along its length and the formation of high-pressure regions at smaller diameters and low-pressure regions at larger diameters. This pressure difference results in liquid flow inside the ligament, hence further reducing the thickness at smaller diameters and increasing it at larger diameters. At later time instances, the smaller diameter regions of the ligament get pinched off, and droplets of larger ligament diameter are formed. This manner of the breakup of droplet ligament is known as the Rayleigh-Plateau mode of instabilities [29,30], as shown in Fig.5(a). The dispersion equation for 1D Rayleigh-plateau instability is as follows[31]:

$$\omega^2 = \frac{\sigma}{\rho_w R_o^3} kR_o \frac{I_1(kR_o)}{I_0(kR_o)}(1-k^2R_o^2) \qquad (3)$$

Where $\omega$ is the growth rate of the instability, $R_o$ is the ligament diameter just before the onset of instability (see Fig. 5(a)), $k$ is the wavenumber, $I_1$ and $I_o$ are the modified

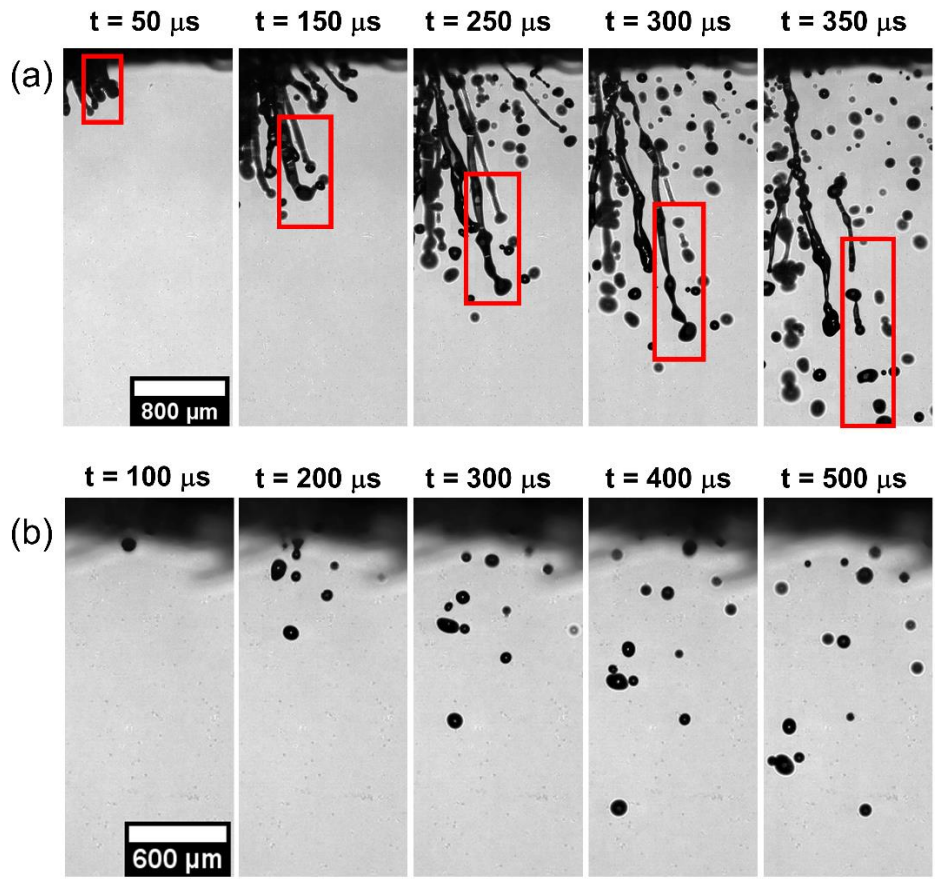

**Figure 4 Zoomed-in images of droplet atomization through mask A.** (a) Impingement on a single-layer mask A is recorded at 20000 fps and "We ~ 880". The impacting droplet extrudes through the mask layer as a cylindrical ligament "(t = 50 μs)" whose length increases over time "(t = 50 - 150 μs)". Unstable waves are formed on the surface of the ligament, which grows in amplitude "(t = 250 - 300 μs)". and leads to its atomization into tiny droplets "(t = 350 μs)" via the Rayleigh plateau instability. (b) Impingement on a double-layer mask. The total number count of the daughter droplets is significantly less, and no ligament formation is observed "(t = 100 - 500 μs)". Similar results are observed for mask B (See Supplementary Fig. S4). Scale bar description is included in the figures.

Bessel functions of first and zero-order, respectively. Equation 3 shows that the instabilities grow over time only if $kR_o < 1$ or $\lambda/R_o > 2\pi$, where $\lambda$ is the instability wavelength. Thus, we plot the growth rate of instability at $0 < kR_o < 1$ for different ligament thicknesses (see Supplementary Fig. S5). The instability with the maximum growth rate occurs at $kR_o \approx 0.697$, which leads to the breakup of the ligament. The breakup time for the ligament can be estimated by inverting it, as follows

$$\tau_b \approx 1/\omega_{max} \tag{4}$$

The daughter droplet size can be obtained by equating surface energies of the ligament and daughter droplets.

$$2\pi R_o L \sigma = N(4\pi R_d^2 \sigma) = \frac{L}{\lambda}(4\pi R_d^2 \sigma) \tag{5}$$

Where L is the length of the ligament, and $R_d$ is the daughter droplet radius. Therefore, on solving we get

$$R_d \approx 2.1 R_o \tag{6}$$

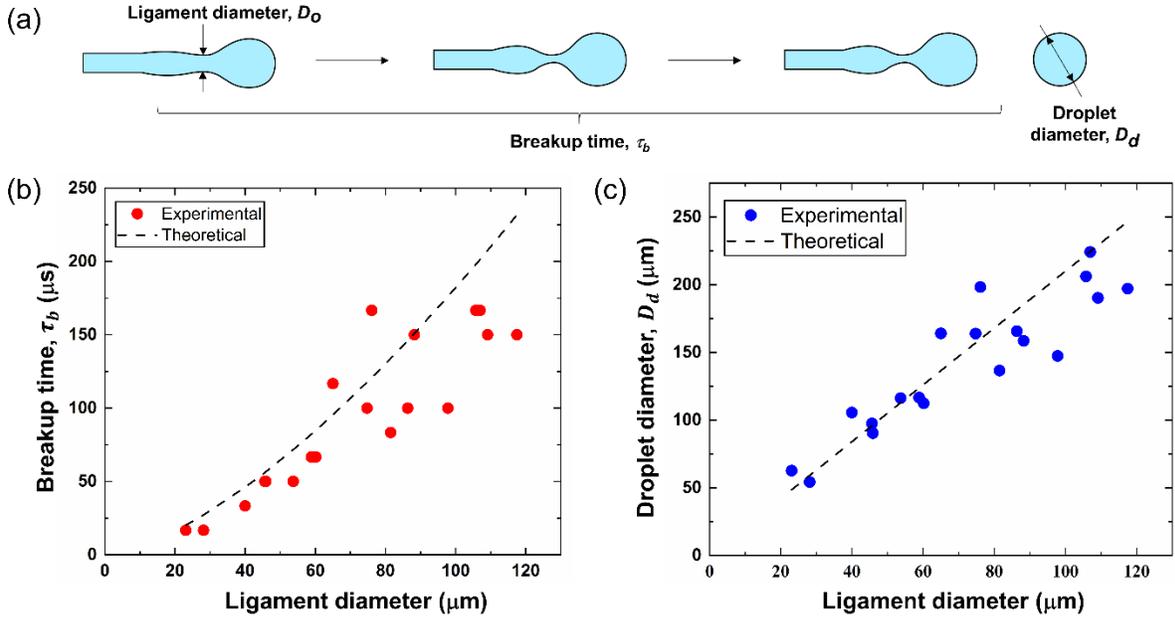

**Figure 5 Breakup time and daughter droplet diameters.** (a) Schematic diagram displaying different stages of a ligament undergoing breakup via Rayleigh-Plateau instability. (b, c) Comparison of experimental and theoretical breakup times and daughter droplet diameters for different ligament sizes.

Due to the hydrodynamic focusing[28] of impacting droplet, the penetrating liquid velocity $\left(\sim \frac{v_i D_i}{\varepsilon}\right)$ is several orders higher than the droplet impact velocity as $\frac{D_i}{\varepsilon} \gg 1$ and hence high recording rate is required for capturing the growth rate of instabilities on ligament surface. Thus, shadowgraphy imaging is done at 60000 fps and pixel resolution of ~ 9 μm/pixel. The results of breakup time and daughter droplet diameters are shown in Fig. 5(b, c), respectively. The uncertainty in measuring ligament thickness and breakup time is of ± 18 μm and ± 16.67 μs, respectively. We have only compared the results for daughter droplets with size > 50 μm because the ligament sizes corresponding to the smaller droplets are beyond the spatial or temporal resolution used in this work. As seen in Fig. 5(b, c), the larger ligament takes a longer time for breakup and vice versa. The growth rate of instabilities is lower for larger ligaments which results in their longer breakup time. Also, the daughter droplet diameter is directly proportional to ligament radius (Eq. 6) therefore we get a larger droplet size for larger ligaments. The theoretical model is found to be in good agreement with experimental data. Figure 4(b) shows a zoomed-in image for a double-layer mask. As discussed earlier, a minimal amount of initial droplet volume penetrates through the mask. No ligament formation is observed in this case due to the presence of the second mask layer.

**Size distribution of atomized droplets**

The probability distribution for the diameter of atomized droplets in mask A and B is shown in Fig. 6. The droplets with a diameter smaller than 100 μm have higher aerosolization tendency[27,32], and these droplets are mentioned as critical droplets in the text hereafter. The atomized droplets are distributed over a size range of 13-288 μm among which 58.48 % and 72.28 % of the droplets are of critical size for single- and double-layer mask A, respectively. Similarly, 64.87 % and 85.82 % of the droplets fall in the critical range for single- and double-layer of mask B, respectively. Although the total number counts of atomized droplets can vary, the probability distribution is similar for all the considered masks indicating that the daughter droplet size range is independent of the mask type used. The initial volume ($V_i$) injected on the mask surface is 123.40 ± 14 nL, out of which 69.88 % and 8.34 % penetrated ($V_p$) the mask, 8.43 % and 2.28 % fell in the critical range ($V_c$) for single- and double-layer of mask A, respectively. While for mask B, 64.3 % and 3.7 % of the initial volume penetrated ($V_p$) the mask, 13.07 % and 1.87 % of which fell in the critical range ($V_c$) for single- and double-layer masks, respectively. Thus, mask B is relatively more effective than mask A in restricting cough droplets for the corresponding number of mask layers. It can also be noticed that the single-layer mask performs poorly not only in restricting the transmission of cough droplets but also atomizes a higher percentage of the initial volume. The double-layer masks perform better in restricting the droplet penetration, but among the droplets which penetrate the mask they are more likely to exist in the critical regime. The average velocity of all the atomized droplets is ~ 1.5 m/s, while the minimum and maximum velocities are ~ 0.12 m/s and ~ 4.2 m/s for both single- and double-layer of mask A and mask B. Thus, these droplets have sufficient momentum (although the momentum is much less compared to the initially ejected droplets) to transmit the viral load to significant distances.

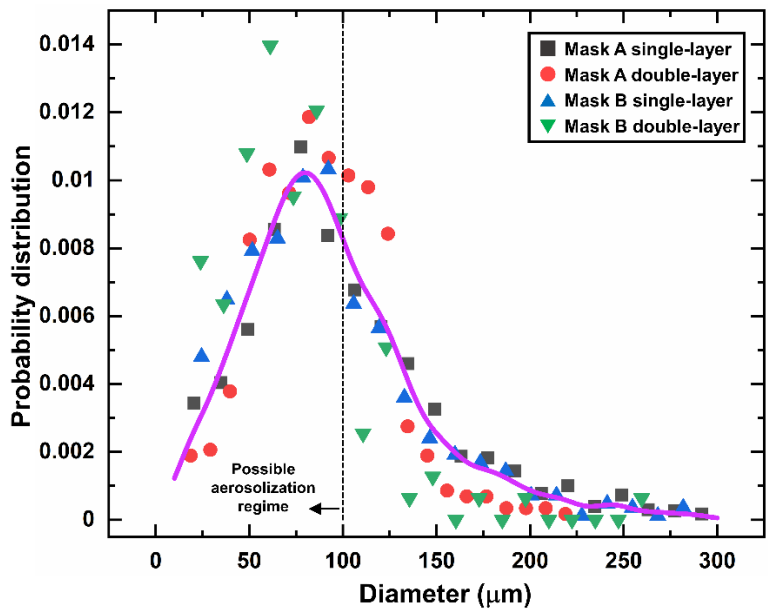

**Fig. 6 Probability distribution of daughter droplets.** A probability distribution is plotted for the daughter droplets that penetrate through single- and double-layer of mask A and mask B. For an impacting droplet of size 620 μm, the most probable size of daughter droplets for both masks A and B (single- and double-layer) falls below 100 μm, which is the regime for possible aerosolization.

**Effectiveness of different masks in trapping viral load**

While the above discussion amply provides insights into efficacy of single-, double- and triple-layer surgical masks in preventing the transport of larger droplets, it does not provide much information on the efficacy in filtering the virus. The cough droplets ejected by an infected person contain virions inside them, and on impacting the mask surface, some of the viral load gets trapped onto its layers, as schematically shown in Fig. 7(a). For finding the efficiency of surgical masks in physically obstructing the viral loads, we prepared a DI water solution loaded with 100 nm fluorescent polymer microspheres (fluoro-max, Thermo fisher scientific) emulating as viruses [33,34] at a concentration of 0.001 wt %. These particles mimic the fluid dynamics of virion laden droplets, although they do not possess the mechanical or chemical properties of virions[33]. The nano-particle loaded droplets are then injected on the face mask, and deposition on the surface is identified from their fluorescent images. Figure 7(b, c) shows the overlayed brightfield and fluorescent microscopy images on impact side and penetration side of the mask, respectively, and images for single-, double-, and triple-layer masks are arranged from left to right. The procedure for image overlaying is described in Supplementary Fig. S6. These images clearly show that some nanoparticles (viruses) get deposited on the mask fibers during the penetration of ligaments. Their quantity will be proportional to the fluorescent signal coming from them. The amount of fluorescent signal is more for a triple-layer mask than single- and double-layer masks, which indicates that it is effective in restricting the viral loads more effectively. On comparing Fig. 7(b, c), a similar fluorescent signal is observed from either side of the face mask. The observation of particle trapping is further verified from the SEM images for a single-layer of mask A (see Fig. 7(d-g)). These images indicate that particles are deposited as a lump (Fig. 7(d, e)) as well as discrete particles (Fig. 7(f, g)) on the surface of the mask. The deposited nano-particles on the mask layer indicates the presence of viruses. This mandates that the user should follow proper disposal methods for handling face masks after utilization.

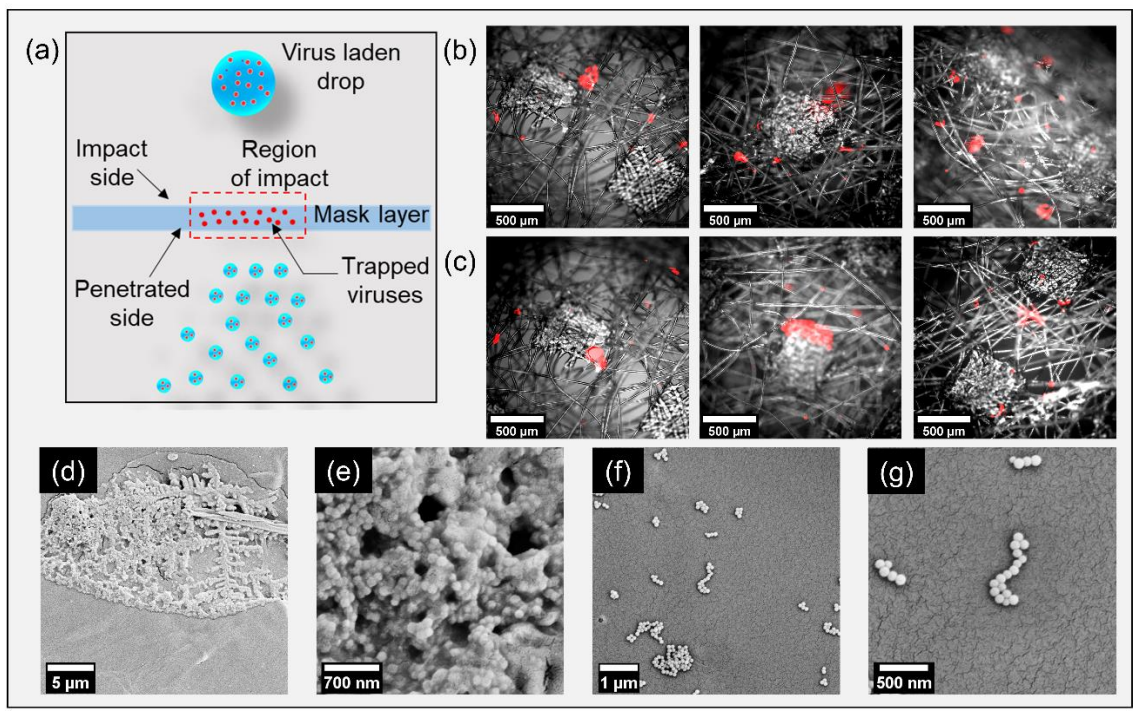

**Fig. 7 Viral load trapping on a mask.** (a) Schematic diagram of viral load getting trapped inside mask layer (b) Overlayed brightfield and fluorescent microscopy images showing trapped particles on the impact side for a single-, double-, and triple-layer of mask A (left to right in order). (c) Overlayed brightfield and fluorescent microscopic images show trapped particles on the penetration side for a single-, double-, and triple-layer of mask A (left to right in order). Similar results are obtained for mask B (See Supplementary Fig. S7). (d) SEM image of particle lump accumulated on mask surface. (e) Zoomed-in image showing coagulation of particles as a lump. (f) SEM image of discrete particles on mask surface (g) Zoomed-in view of particles deposition location. Scale bar description is included in the figures.

## Discussion

The atomization mechanism of large cough droplets impinging on different layered surgical masks is studied in this work. The results of droplet atomization were compared in terms of droplet penetration, size distribution, and volume transmission. Theoretical models for the criteria of droplet penetration, breakup time, and droplet size prediction agree with experimental data. The fluorescent images of particle deposition on the mask layer indicate that some viral loads get trapped onto the mask fibers, thus requiring proper disposal of face masks after their use. Table 2 shows the effectiveness of different surgical masks investigated in this study. Although all masks provide some level of protection, for a droplet of an initial diameter of 620 μm, a single-layer of mask A restricts only 30.12 % of the initial droplet volume and is found to be the least effective among all the tested masks. The double-layer mask performs better comparatively and restricts 91.79 % of the initial droplet volume, but 27.77 % of transmitted droplets fall in the critical droplet diameter regime. Negligible droplet ejection was observed for the triple-layer of mask A. A similar result was obtained for mask B as well. Thus, in the current pandemic situation in which the N95 mask is not easily accessible for the general populace, at least a triple-layer face mask is recommended. This not only restricts the droplet transmission but also prevents the formation of atomized droplets. However, it should be noted that single- and double-layer masks do provide protection in blocking droplet volume (table 2) and is far better than not wearing a mask. It also provides protection against low momentum droplets (table 1) emitted during talking and breathing.

| | Mask type | Initial volume, $V_i$ (nL) | Penetrated volume percentage ($V_p/V_i$) | Percentage Volume in critical regime ($V_c/V_i$) | Number counts of penetrated droplets |
|---|---|---|---|---|---|
| **Mask A** | Single-layer | 123.4 | 69.88% | 8.43% | 100 - 132 |
| | Double-layer | 123.4 | 8.21% | 2.28% | 15 - 42 |
| | Triple- layer | 123.4 | 0.22% | 0.22% | 0 - 1 |
| **Mask B** | Single-layer | 123.4 | 64.3 % | 13.07% | 91-109 |
| | Double-layer | 123.4 | 3.7 % | 1.87% | 11-16 |
| | Triple- layer | 123.4 | 0 % | 0 % | 0 |

**Table 2: Effectiveness of different surgical masks for cough droplets of 620 μm**

We end the exposition by clarifying that any face covering, even the single-layer face masks provide some form resistance against exhalation of respiratory droplets and as such should be used whenever required or mandated by health officials. We also point out that our current investigation only focused on efficacy of single and multi-layer masks in restricting exhaled large respiratory droplets. The assessment of the effectiveness of various masks during the inhalation process requires additional considerations and as such beyond the scope of this study.

## Methods

**High-speed shadowgraphy setup**

The light from a pulsed laser source (Cavitar Cavilux smart UHS) is collimated into a parallel beam using a beam collimator (Thorlabs – BE20M-A). The shadow image of the droplet is captured by a high-speed camera (Photron SA5) coupled with a zoom lens assembly (NAVITAR 6.5x lens, 1.5x objective lens, and 1x adapter tube) at a recording rate of 20000 fps and laser pulse width of 10 ns. A pixel resolution of 11.61 $\mu$m/pixel and 6.64 $\mu$m/pixel is used for zoomed-out and zoomed-in images respectively, with an image resolution of 576 X 624 pixels. The captured images are processed using MATLAB R2019a and Fiji ImageJ softwares.

## References

1.  Liu, Y. *et al.* Aerodynamic analysis of SARS-CoV-2 in two Wuhan hospitals.

*Nature* **582**, 557–560 (2020).

2. World Health Organization. Transmission of SARS-CoV-2: implications for infection prevention precautions. Scientific brief, 09 July 2020. 1–10 (2020).

3. Richard, M. *et al.* Influenza A viruses are transmitted via the air from the nasal respiratory epithelium of ferrets. *Nat. Commun.* **11**, 1–11 (2020).

4. Duguid, J. P. The size and the duration of air-carriage of respiratory droplets and droplet-nuclei. *J. Hyg. (Lond).* **44**, 471–479 (1946).

5. Chaudhuri, S., Basu, S., Kabi, P., Unni, V. R. & Saha, A. Modeling the role of respiratory droplets in Covid-19 type pandemics. *Phys. Fluids* **32**, (2020).

6. Yan, J. *et al.* Infectious virus in exhaled breath of symptomatic seasonal influenza cases from a college community. *Proc. Natl. Acad. Sci. U. S. A.* **115**, 1081–1086 (2018).

7. Chaudhuri, S., Basu, S. & Saha, A. Analyzing the dominant SARS-CoV-2 transmission routes towards an ab initio SEIR model. arXiv:2007.13596 (2020).

8. Chao, C. Y. H. *et al.* Characterization of expiration air jets and droplet size distributions immediately at the mouth opening. *J. Aerosol Sci.* **40**, 122–133 (2009).

9. Robert J. Goodlow, Frederick A. L. Viability and infectivity of microorganisms in experimental airborne infection. *Bacteriol. Rev.* **25**, 182–187 (1961).

10. Alonso, C., Raynor, P. C., Davies, P. R. & Torremorell, M. Concentration, size distribution, and infectivity of airborne particles carrying swine viruses. *PLoS One*

**10**, 1–12 (2015).

11. Kampf, G., Todt, D., Pfaender, S. & Steinmann, E. Persistence of coronaviruses on inanimate surfaces and their inactivation with biocidal agents. *J. Hosp. Infect.* **104**, 246–251 (2020).

12. Casanova, L. M., Jeon, S., Rutala, W. A., Weber, D. J. & Sobsey, M. D. Effects of air temperature and relative humidity on coronavirus survival on surfaces. *Appl. Environ. Microbiol.* **76**, 2712–2717 (2010).

13. van Doremalen, N. *et al.* Aerosol and Surface Stability of SARS-CoV-2 as Compared with SARS-CoV-1. *N. Engl. J. Med.* **382**, 1564–1567 (2020).

14. Leung, N. H. L. *et al.* Respiratory virus shedding in exhaled breath and efficacy of face masks. *Nat. Med.* **26**, 676–680 (2020).

15. Esposito, S., Principi, N., Leung, C. C. & Migliori, G. B. Universal use of face masks for success against COVID-19: evidence and implications for prevention policies. *Eur. Respir. J.* **55**, (2020).

16. Hui, D. S. *et al.* Exhaled Air Dispersion during Coughing with and without Wearing a Surgical or N95 Mask. *PLoS One* **7**, 1–7 (2012).

17. Fischer, E. P. *et al.* Low-cost measurement of facemask efficacy for filtering expelled droplets during speech. *Sci. Adv.* **3083**, 1–11 (2020).

18. Dbouk, T. & Drikakis, D. On respiratory droplets and face masks. *Phys. Fluids* **32**, (2020).

19. Verma, S., Dhanak, M. & Frankenfield, J. Visualizing the effectiveness of face

masks in obstructing respiratory jets. *Phys. Fluids* **32**, (2020).

20. Kähler, C. J. & Hain, R. Fundamental protective mechanisms of face masks against droplet infections. *Journal of Aerosol Science* vol. 148 (2020).

21. MacIntyre, C. R. *et al.* Face mask use and control of respiratory virus transmission in households. *Emerg. Infect. Dis.* **15**, 233–241 (2009).

22. Rodriguez-Palacios, A., Cominelli, F., Basson, A. R., Pizarro, T. T. & Ilic, S. Textile Masks and Surface Covers—A Spray Simulation Method and a "Universal Droplet Reduction Model" Against Respiratory Pandemics. *Front. Med.* **7**, 1–13 (2020).

23. Tang, J. W., Liebner, T. J., Craven, B. A. & Settles, G. S. A schlieren optical study of the human cough with and without wearing masks for aerosol infection control. *J. R. Soc. Interface* **6**, 727–736 (2009).

24. Mittal, R., Ni, R. & Seo, J. H. The flow physics of COVID-19. *J. Fluid Mech.* **894**, 1–14 (2020).

25. Xie, X., Li, Y., Sun, H. & Liu, L. Exhaled droplets due to talking and coughing ( Xie et al, 2009).pdf. *J. R. Soc. Interface* **6**, 703–714 (2009).

26. Zhu, S. W., Kato, S. & Yang, J. H. Study on transport characteristics of saliva droplets produced by coughing in a calm indoor environment. *Build. Environ.* **41**, 1691–1702 (2006).

27. Xie, Li, Y., Chwang, A. T. Y., Ho, P. L. & Seto, W. H. How far droplets can move in indoor environments. *Indoor Air* **17**, 211–2256 (2007).

28. Sahu, R. P., Sinha-Ray, S., Yarin, A. L. & Pourdeyhimi, B. Drop impacts on electrospun nanofiber membranes. *Soft Matter* **8**, 3957–3970 (2012).

29. Hagedorn, J. G., Martys, N. S. & Douglas, J. F. Breakup of a fluid thread in a confined geometry: Droplet-plug transition, perturbation sensitivity, and kinetic stabilization with confinement. *Phys. Rev. E - Stat. Physics, Plasmas, Fluids, Relat. Interdiscip. Top.* **69**, 18 (2004).

30. Lin, S. P. & Reitz, R. D. Drop and spray formation from a liquid jet. *Annu. Rev. Fluid Mech.* **30**, 85–105 (1998).

31. Drazin, P. G. & Reid, W. H. *Hydrodynamic Stability*. (Cambridge University Press, 2004). doi:10.1017/CBO9780511616938.

32. Wells, W. F. On air-borne infection: study II. Droplets and droplet nuclei. *Am. J. Epidemiol.* **20**, 611–618 (1934).

33. Kabi, P., Saha, A., Chaudhuri, S. & Basu, S. Insights on drying and precipitation dynamics of respiratory droplets in the perspective of Covid-19. ARXIV 2008.00934 (2020).

34. Lustig, S. R. *et al.* Effectiveness of Common Fabrics to Block Aqueous Aerosols of Virus-like Nanoparticles. *ACS Nano* **14**, 7651–7658 (2020).